# Point cloud ridge-valley feature enhancement based on position and normal guidance

Jianhui Nie[a*], Zhaochen Zhang[a], Ye Liu[a], Hao Gao[a], Feng Xu[b], Wenkai Shi[a]

a) School of Automation & Artificial Intelligence, Nanjing University of Posts and Telecommunications, Nanjing, China

b) School of Software, Tsinghua University, Beijing, China

**Abstract**—Ridge-valley features are important elements of point clouds, as they contain rich surface information. To recognize these features from point clouds, this paper introduces an extreme point distance (EPD) criterion with scale independence. Compared with traditional methods, the EPD greatly reduces the number of potential feature points and improves the robustness of multiscale feature point recognition. On this basis, a feature enhancement algorithm based on user priori guidance is proposed that adjusts the coordinates of the feature area by solving an objective equation containing the expected position and normal constraints. Since the expected normal can be expressed as a function of neighborhood point coordinates, the above objective equation can be converted into linear sparse equations with enhanced feature positions as variables, and thus, the closed solution can be obtained. In addition, a parameterization method for scattered point clouds based on feature line guidance is proposed, which reduces the number of unknowns by 2/3 and eliminates lateral sliding in the direction perpendicular to feature lines. Finally, the application of the algorithm in multiscale ridge-valley feature recognition, freeform surface feature enhancement and computer-aided design (CAD) workpiece sharp feature restoration verifies its effectiveness.

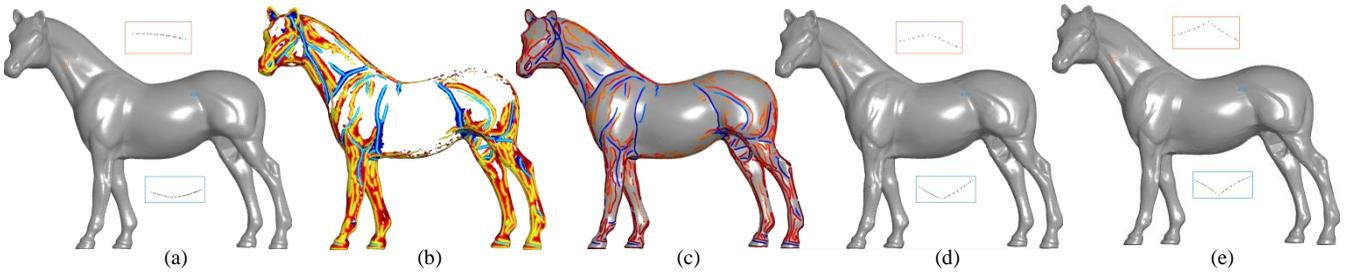

Fig. 1. Given a point cloud (a), our algorithm is able to recognize multiscale ridge-valley features (c) by judging the distance from the current point to the curvature extreme point (b). At the same time, it can effectively enhance features by the position and normal constraints. Different forms of the constraint can produce different enhancement effects, such as maintaining the original features as much as possible (d) or creating sharper features (e).

## 1 INTRODUCTION

Ridge -valley features are important elements in point clouds that can intuitively sketch the basic shape of objects. These features have important applications in fields including surface reconstruction [1], surface editing [2][3], visual perception [4] and multi-perspective data registering [5]. In general, point clouds are obtained by digitizing real objects with vision measuring equipment. During this process, factors such as object surface reflection and sensor quantization error will affect the accuracy of the data acquisition and weaken the sharpness of the original features. In addition, subsequent processing, such as resampling and smooth filtering, may further damage these features. The effective recognition and enhancement of ridge-valley features are important topics in point cloud processing.

To recognize feature points from point clouds, researchers usually employ some indexes, such as the surface variation (SV) [8][9], the smooth shrink index (SSI) [10][11], the curvature [13] [14] [15], to measure the variation in the local surface and then to identify points whose variations are higher than a specific threshold as potential feature points. The real location of feature points can be obtained by iteration refinement or minimum spanning tree (MST) building. However, in this process, it is not easy to find an appropriate truncation threshold, especially in an area with gentle changes. In this case, a large threshold will lead to the incomplete recognition of ridges and valleys, while a small threshold will lead to the excessive recognition of potential feature points, which will have adverse effects on subsequent accurate positioning. How to improve the integrity and accuracy of feature point recognition is a problem that needs further study. In feature enhancement, current research has mainly been focused on the restoration of sharp features in the model, and the general approach has been to optimize the location of features using an anisotropy processing algorithm. Although these algorithms are effective for computer-aided design (CAD) workpiece models composed of standard quadric surfaces, the freeform surface contained in general models can easily cause algorithm failure. Recently, Ming et al. [6] proposed a feature enhancement method for mesh surfaces in gradient domains, which has good generality. However, the algorithm directly amplifies the gradient signal without selection, which makes it sensitive to noise.

In view of the above problems, this paper proposes a new ridge-valley feature enhancement method that is not only suitable for both the CAD model and freeform surface model but also realizes the adjustment of the ridge-valley feature location by restricting the position and normal distribution with prior knowledge. Specifically, the main contributions of this paper are as follows: 1) a simple and effective ridge-valley point recognition method is proposed, which adopts a parametric surface to fit variations of local surface and uses the extreme point distance (EPD) criterion to determine feature points. Compared with traditional methods, the EPD greatly reduces the number of potential feature points and improves the robustness of multiscale feature point recognition. 2) A feature enhancement algorithm based on user priori is proposed. The algorithm adjusts the feature location by solving an objective equation containing the expected position and normal. Through proper transformation, the objective equation can be converted into sparse linear equations to obtain a closed-form solution. 3) A point cloud parameterization method based on feature line guidance is proposed, which reduces the number of unknowns by 2/3 and eliminates the lateral sliding of the point cloud in the direction perpendicular to feature lines. Experiments show that this algorithm can achieve satisfactory results in multiscale ridge-valley feature detection, freeform surface feature enhancement and CAD workpiece sharp feature recovery.

The following sections are arranged as follows: Section 2 reviews related work; section 3 introduces the improved feature point recognition method. Section 4 introduces the proposed feature enhancement algorithm. Section 5 introduces the local point cloud parameterization and its application in solving the enhancement object equation. In section 6, the effectiveness of the algorithm is verified by experiments and comparisons with the mainstream algorithms. Finally, the full text is summarized in section 7.

## 2 RELATED WORK

For a long time, many researchers have studied the problem of ridge valley feature enhancement and obtained rich achievements. These studies mainly focus on feature recognition method and feature enhancement strategy

### 2.1 Feature Recognition

To recognize features points from point clouds, Gumhold et al. [7] introduced the principal component analysis (PCA) method, which uses the eigenvalues of the PCA, to classify potential feature points and find the accurate location of the feature points by constructing an MST. Pauly et al. [8][9] extended the above method and proposed the concept of SV using the ratio of the minimum eigenvalue to the sum of all eigenvalues. Furthermore, the multiscale analysis of local point clouds was conducted to improve the adaptability of the algorithm to noise. Although the SV can well represent the variation degree, it cannot express the convexity of the surface. Therefore, SV demonstrates poor performance for distinguishing features with adjacent spatial distance. In our previous work, the concept of the SSI was proposed [10][11]. The SSI expresses the variation degree of the surface by the shrinking amplitude and determines the convexity of the surface by the included angle between the point normal and shrinking direction; thus, it can effectively distinguish the feature points adjacent in space.

The curvature is also a measurement of the variation of surfaces, which has been widely used in feature recognition. For example, Pang et al. [12] calculated the curvature by quadratic surfaces; Enkhbayar et al. [13] calculated the curvature by fast Fourier transform (FFT) and enhanced the robustness of the computation by low-pass filtering. Kim et al. [14][15] and Weber et al. [16] estimated the curvature using the moving least squares (MLS) algorithm. Kim also obtained the optimized neighbors via a local Voronoi diagram, and similar methods were also adopted by Quentin et al. [17]. Daniels et al. [18] used the robust MLS algorithm to project the potential feature points onto the intersection lines of different surfaces. Overall, the MLS method has good tolerance to noise and therefore has a certain advantage when dealing with point clouds with noise.

For point clouds containing strong noise, statistical analysis is another option to improve computational robustness. For example, Evangelos et al. [19] improved the accuracy of curvature estimation by an M-estimation. Min et al [20] proposed a feature calculation method based on multiscale tensor voting; Weber et al. [21] realized feature line extraction through Gaussian graph clustering.

In addition to measuring the variation of surfaces, it is also very important to choose a reasonable feature recognition criterion. At present, the main solution to this problem is to set a truncation threshold and identify points whose variation is higher than the threshold as feature points. Although simple and easy to implement, the strategy is not applicable to regions with gentle changes. In this case, a large threshold value will lead to the incomplete recognition of ridge-valley points, while a small threshold value will lead to excessive identification of potential features, which will have adverse effects on subsequent accurate positioning. The revision of the existing criteria and an improvement in the integrity of the feature recognition are problems that need further study.

### 2.2 Feature Enhancement

The basic principle of feature enhancement is to adjust the original position of points by anisotropy algorithms. Related algorithms can be divided into three classes: the filter-based method, the projection-based method and the optimization-based method.

The filtering-based method achieves the anisotropy effect by measuring the difference between neighborhood points. The bilateral filtering algorithm [22], which is proposed for image denoising, is a typical representative. For point cloud processing, the weight factor of bilateral filtering is usually constructed by the position and normal difference between neighborhood points. To obtain a better effect, researchers have suggested several normal difference metrics, such as $n_p \cdot n_q$[23], $\|(p-q)\cdot n_p n_q\|$[24], $\|n_p - n_q\|$[25], $n_p \cdot (n_p - n_q)$[26], etc. Moreover, experiments have shown that for point clouds containing strong noise, it is better to first conduct bilateral filtering on the normal of the points and then adjust the position of the features based on the optimized normal [27][28]. Except for bilateral filtering, other edge preserving filters in the image domain can also be extended to point cloud processing, such as mean shift filtering [29], nonlocal mean filtering [30], and guided filtering [31].

The projection-based method considers that the local surface is composed of multiple piecewise continuous surfaces and the core of the algorithm is to select an appropriate projection strategy, which ensures that the point can be projected to the correct surface. For example, the local optimal projection (LOP) [32] achieves a point cloud position adjustment by restricting the offset and uniformity of points. Liao et al. [33] added an anisotropy weight to the LOP and proposed the feature preserved weighted local opti-mal projection (WLOP) algorithm. Preiner et al. [34] proposed a continuous LOP algorithm, which improved the efficiency of the projection. Additionally, many algorithms implement feature reservation through an improved MLS algorithm. For example, based on the M-estimator and MLS algorithm, Mederos et al. [35] provided a new smoothing operator that can preserve salient features. Fleishman et al. [36] introduced a robust MLS technique based on a forward-search paradigm to address noise, outliers and sharp features. Adamson et al. [37] preserved shape features by decomposing the object into cells of different dimensions. Mattei et al. [38] proposed a framework based on moving robust PCA to eliminate outliers and accurately denoise the point cloud.

Inspired by the idea of sparse signal reconstruction, the method based on optimization achieves feature enhancement by solving an objective equation in the sparse solution space. For example, Avron et al. [39] used the L1-sparse criterion to optimize the orientation of the normal first and then restored the sharp features based on the optimized normal. Sun et al. [40] proposed a point cloud denoising algorithm based on L0 minimization, which can recover sharp features while filtering the noise. As the objective equation in the above method is a global expression of point positions, it has better robustness than the local projection method. However, this method is more applicable to CAD data with sharp features and has a poor effect on freeform surfaces.

The main goal of the method described in this section is to filter noise signals in the point cloud. The focus of the algorithm design is how to maintain or recover sharp features in the data during the denoising process. Feature enhancement is only an incidental result. However, during the process of scanning data optimization, surface editing, model reuse and other operations, it is often necessary to modify the data distribution near the features according to the user's prior knowledge. For this reason, this paper studies feature enhancement as an independent problem and proposes a ridge-valley feature enhancement method based on prior guidance. By constructing and solving an objective equation containing position and normal constraints, user-specified coordinate adjustment of the feature area can be achieved. Additionally, the adjusted surface normal can also be an approach to user expectation.

## 3 FEATURE POINT RECOGNITION

To effectively measure the degree of surface variation, this paper adopts the method described in [41] to calculate the curvature of the point cloud. For an arbitrary point $p$ in a given data, a moving least squre surface g(x) as shown in equation (1) is fitted, then the mean curvature of point p can be calculated with equation (2), where n(x) is the gaussian weighted average normal vector of neighbor points, $\nabla g(x)$ is the gradient of g(x) and H(x) is the Hessian matrix of g(x).

$$g(x) = n(x)^T \left( \frac{\partial e(y, n(x))}{\partial y} |_{y=x} \right) \tag{1}$$

$$c = \frac{\nabla g(x) \cdot H(g(x)) \cdot \nabla^T g(x) - \|\nabla g(x)\|^2 \cdot Trace(H)}{\|\nabla g(x)\|^3} \tag{2}$$

After the curvature is calculated, traditional feature recognition algorithms distinguish potential feature points from the point clouds by setting the threshold. However, the feature lines contained in complex models are of various scales, and different truncation thresholds need to be adopted for different feature lines or even different parts of the same feature line, so it is difficult to find an appropriate global parameter.

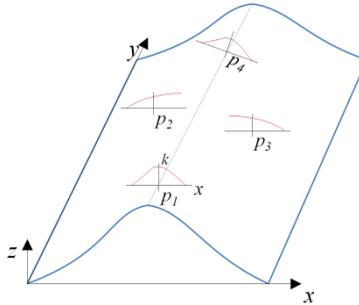

Fig. 2. Schematic diagram of ridge-valley point recognition. In this figure, $p_2$ and $p_3$ are far from the feature line and their curvature changes rise or fall monotonously; $p_1$ and $p_4$ are close to the feature line, and their mean curvature distribution tends to rise first and then fall, and the extreme value is obtained at the feature line. At the same time, this trend is not sensitive to the principal curvature direction errors.

Notably, the smoothness of the quadric surface makes the calculated curvature information change continuously; this is especially true for the average curvature, as it expresses the average variation in the surface and is more consistent with the characteristics of slow and continuous change, which can be well represented by a quadric surface again. Furthermore, as shown in Fig. 2, the ridge-valley points are the curvature extremum points in the maximum principal curvature direction; therefore, the curvature distribution of potential feature points inevitably rises and falls (or falls and rises) near the feature line, and the distance from it to the curvature extreme point is small (as $p_1$, $p_4$ in Fig. 2). In contrast, the curvature distribution of the point cloud far from the ridge-valley points shows a monotonic rise (or fall) trend, and the distance from these points to the curvature extreme point is large (as $p_2$, $p_3$ in Fig. 2). Based on the above analysis, this paper uses a quadric surface to fit the local curvature distribution and recognize ridge-valley points by judging the distance between the current point and the curvature extremum point. The specific steps are as

follows:

(1) Establish an LCS with the current point as the origin, the normal of the current point as the z-axis, and the maximum principal curvature direction as the x-axis. Then, convert the coordinates of the current point and its r neighborhood points to this LCS;

(2) Construct the objective equation shown in equation (2) and solve the parameter $b_0 \sim b_5$, where n is the number of neighborhood points, $x_i$, $y_i$ are the x, y coordinate component of the neighborhood points under the LCS and $c_i$ is the mean curvature of neighborhood points;

$$\arg\min \sum_{i=1}^{n} \left( b_0 x_i^2 + b_1 y_i^2 + b_2 x_i y_i + b_3 x_i + b_4 y_i + b_5 - c_i \right)^2 \qquad (2)$$

(3) Put y = 0 into equation (2) to obtain the curvature curve at the maximum principal curvature direction as $f(x) = b_0 x^2 + b_3 x + b_5$, then calculate the extremum coordinate of $f(x)$ as $x_{max} = -b_3 / (2b_0)$;

(4) Calculate the distance between the current point $p$ and the curvature extremum point as $d = |x_{max}|$. If $d$ is less than the average sampling density $\rho$ of the point cloud, then it can be identified as a feature point; otherwise, $p$ is far away from the real feature point and can be identified as a general point;

(5) Project the extremum point onto the local quadric surface defined by equation (2) to obtain the accurate position of the feature point as ($x_{max}$, 0, $f(x_{max}$,0)).

In the above scheme, a continuous surface with a unique extremum is used to fit the discrete data, and the distance from the current point to the curvature extremum point is taken as the criterion to determine the feature points to ensure that the potential feature points found are distributed in the $\rho$ neighborhood of the real ones. Compared with the method that completely relies on threshold truncation, the method in this paper only uses the internal properties of the surface (e.g., curvature and sample density) to achieve scale independence, especially in the part of the surface with gentle changes. This method greatly reduces the number and distribution range of potential feature points and facilitates the accurate positioning of feature points (Fig. 5). Another advantage of the above algorithm is that the quadric surface fitting has a certain tolerance to curvature noise and has a low requirement for the accuracy of the principal curvature direction. As shown in Fig. 2, there are errors in the principal curvature direction of point p4, but this only changes the amplitude of the curvature distribution in the direction without changing the trend of the curvature distribution, and correct results can still be obtained through the EPD criterion.

After the feature points are obtained, feature lines can be generated by iteration refinement and MST construction [10]. Since the obtained feature points can indicate the accurate position of the feature and can enhance the expression power of the model, this paper adds the feature point $P_f$ to the original point cloud $P$ to form an augmented point cloud $P = P \cup P_f$ and takes $P$ as the operation object in the subsequent processing. At the same time, $P_f$ can also provide more constraint information for feature enhancement, for example, $P_f$ itself can be used as the position guidance point of feature enhancement; the combination of $P_f$ and the direction of the feature lines can provide a basis for determining the neighborhood point's side; by giving $P_f$ a twofold normal vector calculated by neighborhood points on both sides of the feature line, the effect of point cloud rendering and reconstruction can be improved [27].

## 4 FEATURE ENHANCEMENT

Generally, people have some prior knowledge of the ridge-valley feature to be enhanced, which can be described by the position and normal. To integrate prior knowledge into feature enhancement, this paper proposes the objective equation, as shown in equation (3), which is composed of the position constraint term $E_p$ and the normal constraint term $E_n$. Among them, $E_p$ is used to constrain the difference between the enhanced feature position and the user expected position; $E_n$ is used to make the enhanced surface normal conform to user priori; and $\lambda$ is used to balance the importance of both.

$$\arg\min \left( \lambda E_p + (1-\lambda) E_n \right) \qquad (3)$$

The position constraint item $E_p$ has the form of equation (4), where $\tilde{p}_i$ and $\hat{p}_i$ represent the expected and actual positions of the enhanced feature points, respectively.

$$E_p = \sum_{i=1}^{n} \left( \left\| \hat{p}_i - \tilde{p}_i \right\| \right)^2 \qquad (4)$$

The constraint term $E_n$ is described in equation (5), where $\tilde{n}_i$ is the point normal of user expectation, and $\hat{n}_i$ is the normal after feature enhancement.

$$E_n = \sum_{i=1}^{n} \left( \left\| \hat{n}_i - \tilde{n}_i \right\| \right)^2 \qquad (5)$$

In this paper, the point normal is selected as the constraint condition mainly based on the following considerations. On the one hand, the normal vector is a basic attribute of the point clouds, and its calculation method has been widely and deeply studied, so the accuracy of calculation can be easily guaranteed. On the other hand, the normal vector is directly related to point cloud rendering. After the expected normal is assigned to the original point cloud, the outline shape of the enhanced features can be reflected through illumination rendering, which can help users predict the enhanced effects in advance.

Direct nonlinear optimization of objective equation (3) is difficult and inefficient. Fortunately, the point normal can be regarded as the normal vector of the local micro-tangent plane, where the current point and its neighbors are located. Therefore, equation (5) can be described as the expected normal should be as perpendicular as possible to the vector from the enhanced feature point to its neighbors. That is, equation (6), where $k$ is the number of neighborhood points, $\hat{p}_i$ are the coordinates of an enhanced point and $\hat{p}_j$ are the neighborhood points that lay on the same side of the feature line with $\hat{p}_i$.

$$E_n = \arg\min \sum_{i=1}^{n} \sum_{j=1, j\neq i}^{k} \left( \left\| \tilde{n}_i \cdot \left( \hat{p}_i - \hat{p}_j \right) \right\| \right)^2 \qquad (6)$$

In combination with equations (3), (4) and (6), the final objective equation can be obtained as follows:

$$\arg\min \left( \sum_{i=1}^{n} \left( \lambda \left( \left\| \hat{p}_i - \tilde{p}_i \right\| \right)^2 + (1-\lambda) \sum_{j=1, j\neq i}^{k} \left( \left\| \tilde{n}_i \cdot \left( \hat{p}_i - \hat{p}_j \right) \right\| \right)^2 \right) \right) \qquad (7)$$

In equation (7), each feature point can give three equations derived from the position constraint and $k/2$ equations derived from the normal constraint (only neighbor points of one side are used). Therefore, the number of equations is $1+k/6$ times the number of unknowns, which belongs to the over-determinate equation. Equation (7) is a linear combination of positions (unknowns) to be solved. Therefore, the closed solution can be obtained in the sense of linear least squares.

In fact, equation (7) provides a general framework for feature enhancement. Different effects can be achieved by assigning different forms of position and normal constraint term. Typically, this paper considers the following three cases: **(1) conformal enhancement**: the user gives the expected feature enhancement amplitude $\delta$ and requests to maintain the local property of the surface as much as possible. In this case, we move feature points $P_f$ in the augmented point cloud $P'$ along its normal with $\delta$ as the seed point of the expected position and keep the other points in $P'$ unchanged. Through the guidance of the seed points and the least squares criterion, feature enhancement can be dispersed to every point in the point cloud to achieve an adjustment of feature position. Moreover, the original normal vector of $P'$ is taken as the expected normal in the adjustment process to maintain the local property of the point cloud to the greatest extent. (2) **Sharp feature recovery**: In the process of point cloud acquisition, the image quality of measuring equipment is easily affected by the ambient light or the reflection of the object surface, which makes sharp features smooth and blurred. In that case, feature normal can be considered the same as the local plane on the left and right sides of the feature line [26] [28] [32]. To infer the destroyed feature normal, we divide the local point into two subsets, the left and the right, with feature line direction. Then, calculate the average position and normal of the point on each side to generate the left and the right mean plane. Finally, the projection of feature nodes at the intersection line formed by the left and right mean plane is set as the expected position, and the linear weighted average of the mean plane normal and original normal by distance to feature lines is set as the expected normal. The enhancement amplitude can be determined in two ways: one is the user-set method, which is the same as the conformal mode, and the other is automatic conjectural by project feature points to the intersection line left and right mean plane. (3) **CAD feature** restoration: Different from freeform surfaces, elements in CAD models usually have a specific angular relationship, which can be well defined by the left-right mean plane. Therefore, after calculating the left-right mean plane, the expected position can be set as the projection of the current point on the mean planes, which can further increase the reliability of position constraints. As the point cloud close to the feature line is blurred, it will affect the accuracy of the mean plane calculation. Therefore, in the following experiments, we ignore those point clouds; that is, when determining the mean plane, we only use points whose distance to the feature lines are in the range of $3\rho \sim 5\rho$.

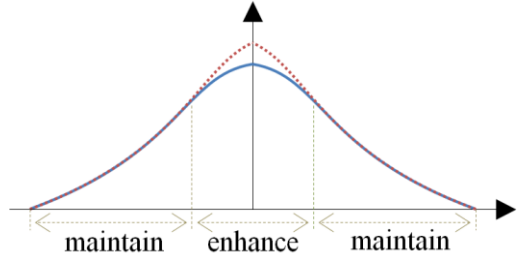

Fig. 3. Enhance region and maintain region.

Equation (7) can be used to modify the point cloud of the feature areas, but it is equally important to keep the data of nonfeature areas unchanged. For this reason, as shown in Fig. 3, the neighborhood of the feature line is divided into two parts: the enhance region and the maintain region. For the points in the enhance region, constraint conditions are given according to equation (7), while for the points in the maintain region, the position and normal are kept unchanged to achieve a smooth transition between the enhanced point cloud and the original point cloud. In the subsequent experiments, unless otherwise specified, the radius of the enhancement region is set as $3\delta$, while the maintain region is set as the annular area with an inner diameter of $3\delta$ and an outer diameter of $5\delta$.

## 5 FEATURE REGION PARAMETERIZATION

Fig. 9(a) demonstrates that if the drift direction of points is not constrained during the enhancement, then the resulting point cloud will slide laterally in the direction perpendicular to the feature line, resulting in missing data in the feature region and affecting the enhancement effect. A direct way to overcome the problems is adding extra constraints to equation (7) so that the movement of points can be carried out in the same direction. However, this addition will increase the number of equations and reduce the efficiency of solving the equations. However, determining the weight of the new constraint to achieve an appropriate balance between feature enhancement and topological maintenance must also be solved.

Notably, the local surface near the blurred feature to be enhanced always changes continuously. Therefore, if the point normal is parallel to the z-axis, then its neighbors can form a one-to-one mapping with the projection in the XOY plane, which can keep the original topological structure unchanged. In this case, only the z coordinates of the points need to be updated (keep the x and y coordinates unchanged) to achieve feature enhancement. Since all points move along the z-axis, this method can effectively overcome the problem of lateral sliding. At the same time, the number of unknowns will be reduced to 1/3 of the original, so the computational efficiency will be greatly improved. To rotate all feature points to the above ideal state, for any node $v_i = (v_{xi}, v_{xi}, v_{zi})$ of the feature line and its affiliated points (points take $v_i$ as the nearest feature point), the following parameterization method is proposed:

(1) Construct an LCS with $(0, 0, l)$ as the origin, feature line direction $\pmb{dir}_i$ at $v_i$ as x+, and normal vector $\pmb{n}_i$ of $v_i$ as z+, where $l$ is the cumulative length from the first node of the feature line to the current node $v_i$;

(2) Calculate the rigid transformation $\pmb{T}_i = (\pmb{R}_i, \pmb{t}_i)$ that translates current points to the LCS; that is, $v_i' = \pmb{R}_i (v_i - \pmb{t}_i)$, where $\pmb{t}_i = [v_{xi}, v_{xi}, v_{zi} - l]'$ and $\pmb{R}_i = [\pmb{dir}_i, \pmb{n}_i, \pmb{dir}_i \times \pmb{n}_i]$;

(3) Transform the affiliated points of node $v_i$ to the LCS using $\pmb{T}_i$.

The key of the above strategy is to calculate the node transformation $\pmb{T}_i$. Due to the inherent property of a rigid transformation, errors of the rotating components will be enlarged as the distance from the origin increases. Therefore, when they are far from the feature line, it is difficult to maintain the original smoothness of the affiliated points between different feature lines after parameterization (Fig. 10(c)). To overcome this problem, this paper rotates the affiliated points around the x-axis of the LCS using the objective equation shown in equation (8), where m and n are the number of feature line nodes and the number of neighborhood points, respectively. $\Delta\theta_i$ is the rotation increment in node $v_i$, $\pmb{n}_i$ is the normal vector of the neighborhood point under LCS, $p_i$ and $p_j$ are the coordinates of the current neighborhood point and its k-nearest neighbors. Subsequent experiments show that equation (8) can eliminate the influence brought by the rotation error and make the enhanced point cloud maintain its original smoothness.

$$\arg\min\left(\sum_{i=1}^{m}\Delta\theta_i^2 + \sum_{i=1}^{n}\sum_{j=i-1,j\ne i}^{i+1}\left(\left\|n_i\cdot\left(p_i^{'}-p_j^{'}\right)\right\|\right)^2\right) \tag{8}$$

When the rotation increment $\Delta\theta$ is small, it satisfies $\sin\Delta\theta \approx \Delta\theta$, $\cos\Delta\theta \approx 1$. Therefore, the coordinates of affiliated points before and after the rotation satisfy the linear transformation relationship described in equation (9). As a consequence, equation (8) becomes a linear equation for the rotation increment $\Delta\theta$, and the closed solution can be obtained.

$$p' = \begin{bmatrix} 1 & 0 & 0 \\ 0 & \cos\Delta\theta & -\sin\Delta\theta \\ 0 & \sin\Delta\theta & \cos\Delta\theta \end{bmatrix} p \Leftrightarrow \begin{cases} p_x^{'} = p_x \\ p_y^{'} = p_y - p_z\Delta\theta \\ p_z^{'} = p_y\Delta\theta + p_z \end{cases} \tag{9}$$

# 6 Experiments

## 6.1 Feature Recognition

During the feature recognition stage, the parameters to be specified are mainly the neighborhood radii used in local data fitting by equations (1) and (2). Experiments show that good results can be achieved by setting the parameters as 3 times the average sampling density of the point cloud. Therefore, in the subsequent experiments, we take $r = 3\rho$ as the default.

Notably, the method of fitting local data by a quadric surface is only valid when there is at most one ridge-valley feature in the fitting area. Therefore, as shown in Fig. 4, if there are feature lines close to each other in the data, the points of multiple feature lines may be included in the neighborhood searching (the red part and the yellow part in the figure), resulting in a failure to represent the distribution by a quadric surface. To solve the problem, initial neighborhood points are screened as follows: First, the points whose convexity is different from the current point are removed (the blue area in the figure). Then, the region growing method described in [10] is adopted to filter out the neighbors that are not directly connected with the current point (the yellow part in the figure). After the above selection, only the neighborhood points of the current feature line can be reserved in the search results, which satisfy the single extremum property of the quadric surface.

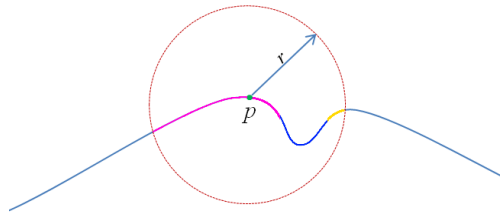

Fig. 4. Selection of neighborhood points for curvature fitting. (A large search radius results in the inclusion of the data of adjacent ridges and valleys into the search results. After filtering by convexity and connectivity, only neighborhood points of the current feature line are retained).

To verify the effectiveness of the feature recognition algorithm in this paper, the continuous surface, expressed as $z = 2\sin x(1 + y/10)$, is uniformly sampled in a rectangular region with points (0,0) and $(2\pi, \pi)$ as the diagonal vertices and generates a discrete point cloud, as shown in Fig. 5(a). According to the surface equation, the model has two feature lines parallel to the y axis at $x = \pi/2$ and $3\pi/2$, and the feature scale increases with increasing y coordinate. Due to the scale changes in the different parts of a single feature line and between different feature lines, it is difficult to obtain ideal results when using the global truncation threshold method. If the minimum curvature of the feature points is used, then the integrity of recognition can be guaranteed. However, too many potential feature points will be identified, which will cause errors in the subsequent accurate positioning (as shown in Fig. 5(c)). Using a larger threshold can reduce the number of potential feature points, but some feature points will be missed detection, affecting the integrity of recognition. Fig. 5(d) shows the heat map of EPD calculated by the method described in section 3. The result can accurately reflect the distance from the current point to the real feature point. The feature points recognized by the distance heat map are shown in Fig. 5 (e). The results reveal that, under the joint action of the distance criterion and the extremum point projection strategy, the feature lines of all scales are correctly recognized, and the recognized potential feature points are located on the feature lines exactly. After connecting them successively, smooth and complete feature lines can be obtained.

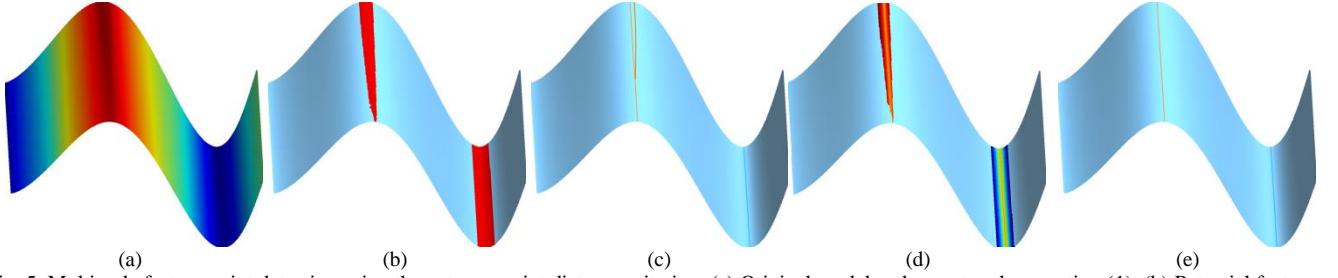

Fig. 5. Multiscale feature point detection using the extreme point distance criterion. (a) Original model and curvature by equation (1). (b) Potential feature points by threshold truncation. (c) Feature line by thinning all potential feature points; excessive points cause bifurcation of the identified feature lines. (d) Heat map of the distance to curvature extreme point. (e) Potential feature points filtered by the extreme point distance criterion and final feature lines.

Fig. 6 is an example of feature recognition from a real scanned CAD workpiece. This example shows that the calculated results of equation (1) can well reflect the undulation degree of the surface. Meanwhile, the distance from the current point to the curvature extreme point is also well measured. After setting the point cloud sampling density $\rho$ as the distance threshold, the algorithm can recognize all the ridge-valley lines in the model. In addition, after adding feature points to the overall point cloud, clear divisions can be formed between different patches, which is helpful to surface segmentation. Fig. 7 shows the result of feature line recognition from a freeform surface. The complexity of the frog body determines that there must be multiple scale features in the model. Nevertheless, the algorithm realizes feature recognition of the different scales with global uniform parameters. Fig. 8 is a more challenging example. There are many feature lines close to each other. To ensure the recognition effect, we filter neighborhood points according to the method shown in Fig. 4. That is, when performing local data fitting, only neighborhood points with the same curvature sign and directly connected with the current point are used, which effectively avoids the error caused by the data confusion of different feature lines.

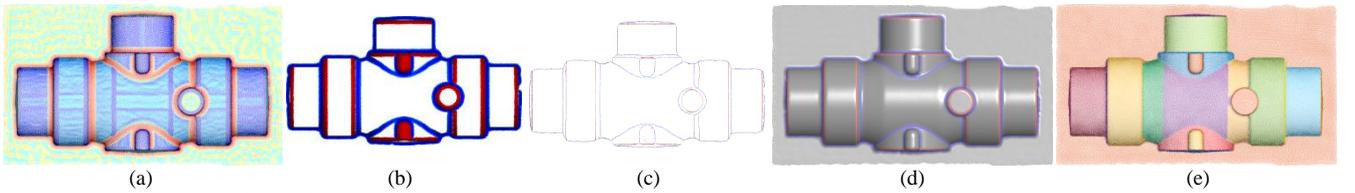

Fig. 6. Feature recognition from a CAD workpiece. (a) Original model and curvature by equation (1). (b) Potential feature points by the curvature threshold. (c) Potential feature points filtered by the EPD criterion. (d) Feature lines. (e) Surface segmentation based on extracted feature lines.

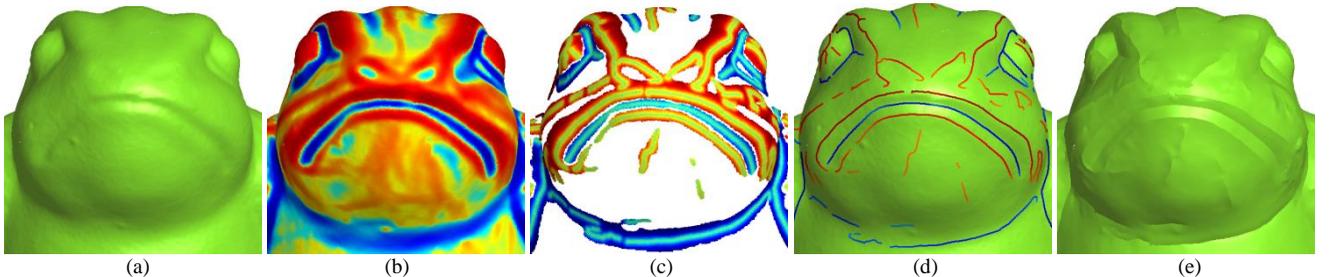

Fig. 7. Feature detection and enhancement of the frog model. (a) Original model. (b) Curvature by equation (1). (c) Potential feature points by threshold truncation and the heat map of distance to the curvature extreme point. (d) Feature lines according to our EPD criterion. (e) Feature enhancement with conjectural amplitude under a sharp recovery mode.

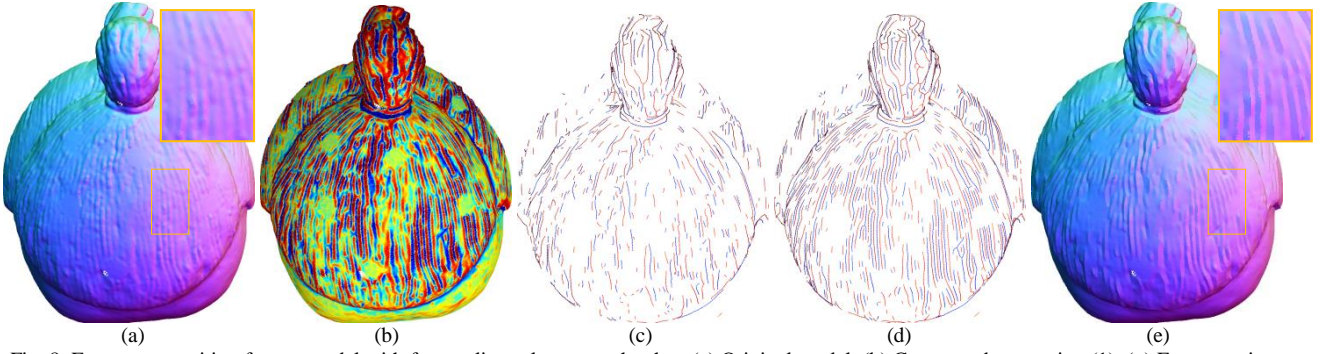

Fig. 8. Feature recognition from a model with feature lines close to each other. (a) Original model. (b) Curvature by equation (1). (c) Feature points recognized by all neighbor points. (d) Feature points recognized by selected neighbor points. (e) Feature sharpens with magnitudes of $3\rho$ under the sharp recovery mode.

## 6.2 Feature Enhancement

**Experiments based on synthetic data**: In the experiment shown in Fig. 9, we generated an asymmetric convex surface whose expression is $z = \sin\left(\sqrt{x}\right)$ and processed its features. The experiment first verifies the importance of the constraining point movement direction in the process of enhancement: As shown in Fig. 9(a), if points are allowed to move freely in space, the point near the feature line will slide laterally, leading to the loss of data at the feature region. In contrast, if we limit the movement direction of points to be parallel to the z-axis, as shown in Fig. 9(b) and (c), then the result maintains the original sampling density and obtains an ideal enhancement effect. Fig. 9 also shows the enhancement results by the conformal mode, sharp feature recovery mode and CAD mode, respectively. As shown in the figure, the results show different characteristics under different modes. For example, in the conformal mode, the algorithm directly takes the normal of the original point cloud as the expected value. By doing so, the algorithm can maintain the gradient information of the original surface as much as possible while enhancing the feature intensity to finally achieve the conformal enhancement effect. In the sharp feature recovery mode, the algorithm uses the weighted average of the mean plane normal and the original normal according to the distance to the feature points as the expected normal. As a result, the position close to the feature line is more in line with the mean plane normal, and the position far from the feature line is more in line with the normal of the original features. Finally, under the dual constraints of the expected position and normal, the algorithm can restore a sharp feature at the feature line and achieve a smooth transition between the enhanced point cloud and the original point cloud. In the CAD mode, the algorithm directly projects the local point cloud onto the mean plane to infer the real positions of the original surfaces. Therefore, the enhanced points completely present plane characteristics near the feature line. We also carried out multiple experiments with different enhancement amplitudes under the conformal mode and sharp feature restoration mode. The results reveal that our algorithm can enhance features with different amplitudes while effectively maintaining the consistency of the result normal with the expected value.

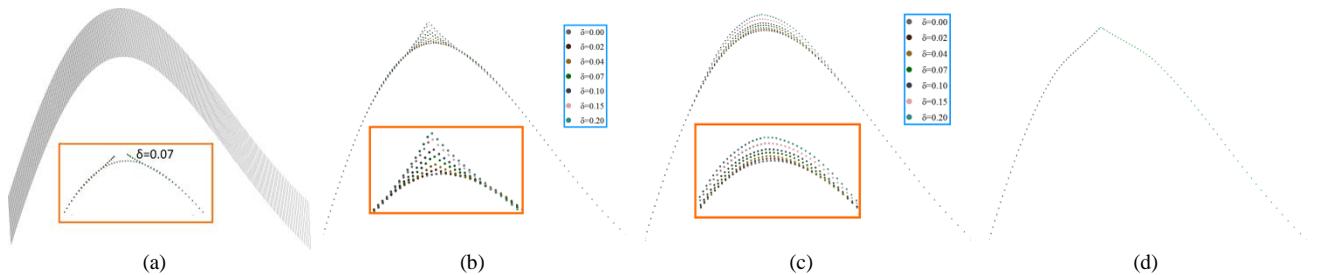

Fig. 9. Feature enhancement of an asymmetric convex surface. (a) Original model, local zoom reveals that the free movement of points will lead to data missing near the feature line. (b) Feature enhancement under the conformal mode. (c) Feature enhancement under the sharp feature recovery mode. (d) Feature enhancement under the CAD mode.

The experiment in Fig. 10 confirms the importance of rotation optimization in parameterization. Due to the influence of noise, there may be errors in the normal calculation of the feature line nodes. If these node noarmals are directly used, as shown in Fig. 10(c), such rotation matrix errors will be transmitted to the parameterized results, causing rotation dislocation among the affiliated points of different nodes and finally leading to strip folds, as shown in Fig. 10(e). Fig. 10(d) shows the parameterized point cloud after the rotational optimization using equation (8). The optimized results well maintained the smoothness of the original point cloud. As shown in Fig. 10(f), the fold disappears in the enhancement result, which indicates that rotation optimization can effectively overcome the influence of normal errors and improve the parameterization effect.

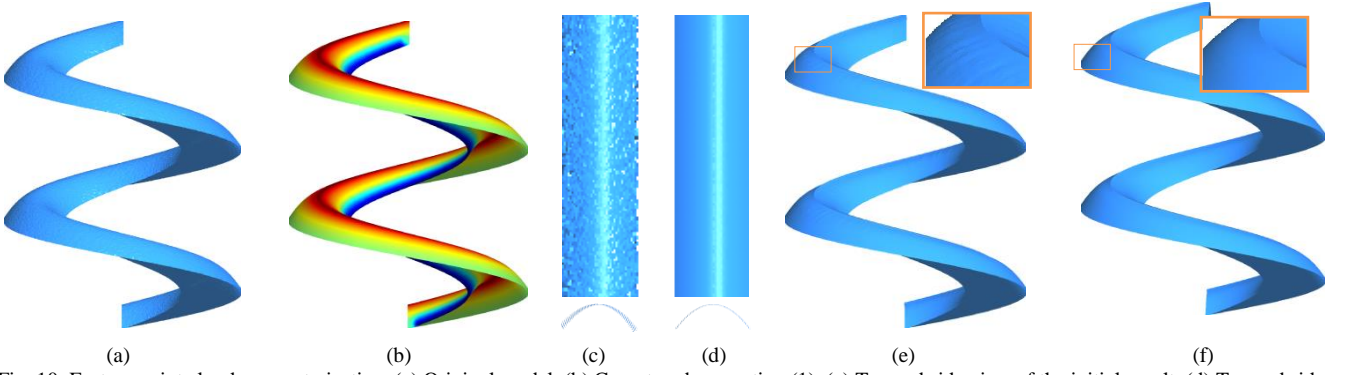

Fig. 10. Feature point cloud parameterization. (a) Original model. (b) Curvature by equation (1). (c) Top and side view of the initial result. (d) Top and side view of the refined result. (e) Feature enhancement using the initial result. (f) Feature enhancement using the refined result.

**CAD Sharp Feature Recovery**: Fig. 11 displays the experimental result of the sharp feature recovery of the CAD model. Since the exact location of the feature line is obtained in the preorder experiments, the point clouds close to the feature line are excluded when calculating the expected position and normal vector. Therefore, the calculation can well reflect the local plane where the real feature is so that the algorithm can satisfactorily recover the sharp features in the model. As seen from the local zooming view, the points normal maintains a gentle change in the nonfeature area, while a mutation appears near the feature line, which indicates that the algorithm can correctly distinguish different surface patches. After adding the feature points to the overall point cloud, the intersection line of the different surface patches is further clarified, which can provide more explicit guidance information for later processing and improve the detailed expression of the surface. Except for the method of this paper, the bilateral filtering algorithm can also recover sharp features using the difference between neighborhood points. However, the bilateral filtering algorithm can work only if the normal variation within a neighbor radius is greater than a certain threshold. As the feature scale in the same model often changes, it is difficult to find a uniform neighborhood radius and weight parameter for processing, especially for a feature with gentle changes. If the filter parameters are not properly selected, then excessive smoothness, as shown in Fig. 11(b), will occur. At the same time, due to the discrete characteristics of the point cloud, the restored features often fail to reflect the smoothness of the original feature line. Edge aware upsampling (EAR) [28] is another algorithm that can restore sharp features in the model. This method first optimizes the normal based on the bilateral filtering algorithm and then adjusts the position of features based on the optimized normal. Due to the lateral sliding of the point cloud in the local plane, the data around the feature line will be missing. Therefore, the method requires a point generation algorithm to add new data to ensure the integrity of the feature. Although the added points can optimize the distribution of the point cloud near the feature line to some extent, it cannot guarantee that the points fall exactly on the feature line due to the lack of guidance from real feature positions (Fig. 11(c)). In contrast, the algorithm in this paper directly adds feature line nodes to the point cloud, which can restore feature information to a greater extent to obtain more optimized results (Fig. 11(d)).

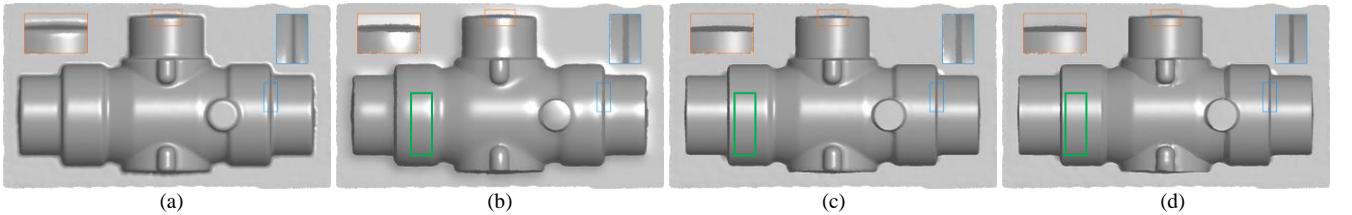

Fig. 11. Feature enhancement of the CAD-like model. (a) Original model. (b) Bilateral. (c) EAR. (d) Our method.

**Freeform Surface Feature Enhancement**: Fig. 8(d) and Fig. 12 show the results of feature enhancement of a freeform surface. Since there are a large number of feature lines close to each other in the model, if the enhancement and maintenance strategy in Fig. 3 are directly adopted, the influence areas of different feature lines will overlap with each other, leading to solution errors. To overcome this problem, we first assign each point in the feature region to its nearest feature point, and the minimum bounding radius r of the affiliated points of each feature point was calculated. Then, the enhancement amplitude is set as r/5, and the enhancement area is set as r/3. Experiments show that the strategy above can adaptively determine the enhancement amplitude and prevent the occurrence of errors in the areas where the feature lines overlap effectively. Furthermore, the parameterization of the feature region enables the algorithm to effectively maintain the sampling density of the original point cloud. Notably, the EAR algorithm [28] also has the ability to enhance freeform surface features. However, the enhancement amplitude is reversed determined by the optimized normal, rather than a variable that can be set artificially. Therefore, the enhancement of an arbitrary amplitude cannot be achieved. Additionally, as mentioned before, the algorithm cannot guarantee that the newly added points fall on the feature line, so the detailed performance of the enhanced point cloud is inferior to our algorithm. Recently, in reference [6], an enhancement algorithm was proposed by amplifying the original signal in the gradient domain. Although the enhancement amplitude can be set arbitrarily, it does not select and filter the amplified signal, and the small noise contained in the original signal is easily amplified, resulting in insufficient smoothness of the enhanced surface, especially the point cloud near the ridge and valley features (Fig. 12(e)). Additionally, the algorithm needs to reconstruct the point cloud to mesh surfaces to be applicable and can only realize feature enhancement of the conformal mode and cannot modify features based on prior knowledge, such as sharp feature restoration of the freeform surface and feature line processing of the CAD model.

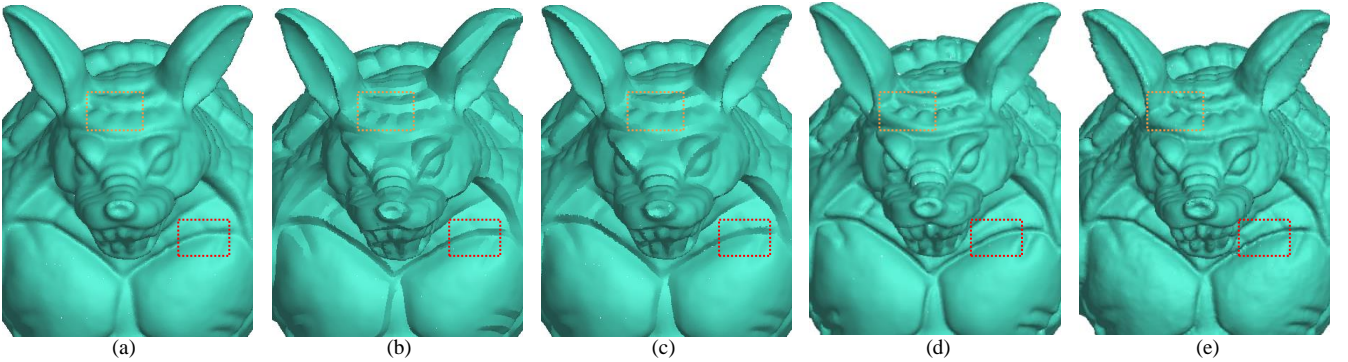

Fig. 12. Feature enhancement of the Armadillo model. (a) Original model. (b) Our method (sharpen enhancement). (c) EAR. (d) Our method (conformal enhancement). (e) Gradient domain enhancement (point cloud is meshed by the ball pivoting algorithm [42]).

For noise data, this paper first uses a relatively large neighborhood radius to calculate the normal point cloud and then uses equation (1) to calculate the curvature. Although a large neighborhood radius will cause the normal near the feature line to be smooth, which is not in line with the true normal distribution of the original surface, it can precisely filter out the influence of noise, making the curvature obtain an extreme value at the feature line, so that the algorithm can obtain correct results. For the algorithm in this paper, the key to success is not the absolute accuracy of the curvature, but the calculated results can correctly reflect the trend of curvature on both sides of the feature line. The above feature makes the algorithm insensitive to the calculation error of the principal curvature direction. In the experiment, we randomly deflect the principle curvature direction of the model in Fig. 5 by $[\pi/6, -\pi/6]$, and the identification result is not affected. The reason is that the deflection of the principle curvature direction only changes the amplitude of the curvature distribution but does not change the trend of the curvature distribution. Therefore, the correct result can still be obtained through the extremum point criterion. In the feature enhancement stage, the expected position and normal of the freeform surfaces and CAD workpiece are calculated by the mean value of the neighborhood points. Therefore, this approach has a greater tolerance to noise. For conformal enhancement, the expected position is obtained by the deviation of the current point along the normal direction of its nearest feature point. To overcome the conduction of the noise, we first calculate the mean point of the neighborhood and project the calculated results to the line where the normal vector is located and then obtain the expected position based on the updated points. In addition, constraints in equation (7) are based on the L2 criteria, making the algorithm tend to homogenize the error between the calculation results and expected values to help overcome the influence of noise. Experiments in Fig. 13~Fig. 14 verify the effectiveness of the above strategies.

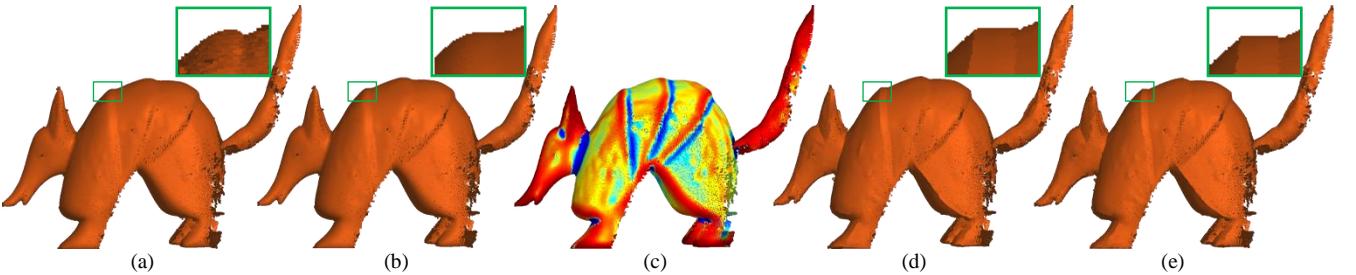

Fig. 13. Feature point detection from a noise model. (a) Original model. (b) Smoothed result using the neighbor radius of 6ρ. (c) Curvature by equation (1). (d) Feature enhancement result of our algorithm with a conjectural amplitude. (e) Feature enhancement result of EAR.

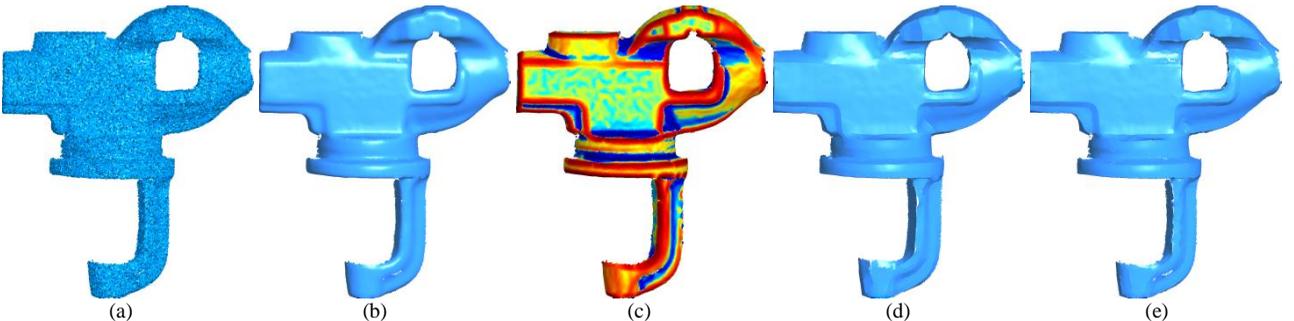

Fig. 14. Feature enhancement of the noise vise model. (a) Original model. (b) Smoothed result using the neighbor radius of 6ρ. (c) Curvature by equation (1). (d) Feature enhancement result of our algorithm with the CAD mode. (e) Feature enhancement result of EAR.

In addition to noise, our algorithm has a good tolerance for nonuniform sampling. Fig. 15(a) shows a point cloud derived from linear structured light scanning. Since the motion interval of the measure device at adjacent moments is much larger than the point density of a single scanning line, the point cloud shows obvious nonuniform characteristics. Nevertheless, experiments in Fig. 15(b)-(d) show that such nonuniform data can still provide enough support for the fitting of the local curvature distribution to complete the recognition of feature lines. In the feature enhancement stage, the normal constraint constructed based on local neighbors can build an interconnected network so that the constraint conditions can be transmitted to each data point to finally overcome the influence of nonuniform sampling. The experimental results in Fig. 15(d), (e) verify the correctness of the above analysis, indicating that the algorithm can extract feature lines and recover sharp features from nonuniform sampled models.

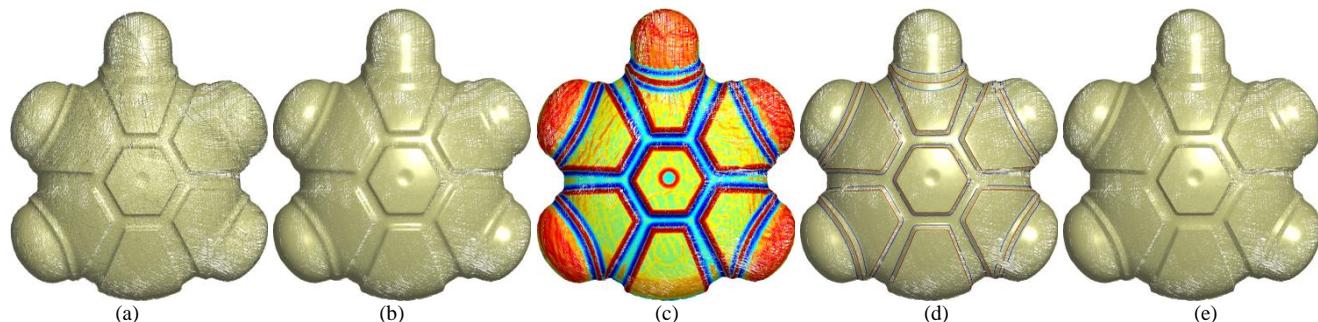

Fig. 15. Feature enhancement of a nonuniform sampled model. (a) Original model. (b) Smoothed result using the neighbor radius of $6\rho$. (c) Curvature by equation (1). (d) Detected feature lines. (e) Feature enhancement result of our algorithm with the CAD mode.

# 7 CONCLUSIONS

In this paper, the EPD criterion for ridge and valley feature point recognition is proposed. By fitting a local curvature distribution with a quadric surface and judging the distance between the current point and the curvature extremum point, multiscale feature points are recognized at once. On this basis, we construct a general framework of feature enhancement with expected position and normal as constraints and give three specific feature enhancement methods in typical cases. In addition, we present a local point cloud parameterization method based on feature line guidance, which reduces the number of unknowns by 2/3 and eliminates the lateral sliding of the point cloud in the direction perpendicular to feature lines. The verification experiment and the comparison with mainstream algorithms show that our algorithm can achieve ideal results in aspects of conformal enhancement and sharp feature recovery and has a good tolerance to noise and nonuniform sampling.

Currently, only three cases of feature enhancement are considered in this paper. In fact, equation (1) has potential application value in point cloud denoising, surface editing and other aspects. In the following work, we will study the above problems. At the same time, only the nearest distance principle is adopted when defining the affiliated points, which leads to the independent enhancement of each feature line. However, when there are feature lines close to each other, it is more reasonable to comprehensively consider the overlapping influence of the neighbor points. Finally, under the condition of strong noise or outliers, if $L0$ or $L1$ constrained projection methods can be used to preprocess the data in advance, then the processing effect of the algorithm can also be further improved.

## ACKNOWLEDGMENT

We would like to thank all the reviewers for their valuable comments. We also thank the authors of [28], AIM Shape Repository and the Stanford Repository for providing the models used in this paper. This work was supported in part by grants from National Nature Science Foundation of China (No. 61802204, 61571236, 61931012, 61602255).